\documentclass[onecolumn,preprint,showpacs,preprintnumbers,amsmath,amssymb,superscriptaddress,floatfix]{revtex4-1}

\usepackage[nottoc,numbib]{tocbibind}

\usepackage[colorlinks=true, citecolor=black, linkcolor=black, urlcolor=black]{hyperref}

\usepackage[all]{hypcap}
\usepackage{amsmath}
\usepackage{enumerate,bm}
\usepackage{graphicx}
\usepackage{grffile} 
\usepackage{color}
\usepackage{esint}
\usepackage{textpos}
\usepackage[usenames,dvipsnames]{xcolor}
\usepackage{subfigure}
\usepackage{flushend}
\usepackage{tabularx}
\usepackage{bbold}

\usepackage{amssymb}
\usepackage{float}
\usepackage{subfigure}
\usepackage{ulem} 
\usepackage[parfill]{parskip} 

\setcitestyle{round}

\newcommand{\vk}{{\vec{k}}}
\newcommand{\vq}{{\vec{q}}}

\newcommand{\ut}[1]{\mathrm{\; #1}}

\begin{document}

\author{M.~A.~Sentef}
\email[]{michael.sentef@mpsd.mpg.de}
\affiliation{Max Planck Institute for the Structure and Dynamics of Matter,
Center for Free Electron Laser Science, 22761 Hamburg, Germany}

\author{M.~Ruggenthaler}
\affiliation{Max Planck Institute for the Structure and Dynamics of Matter,
Center for Free Electron Laser Science, 22761 Hamburg, Germany}

\author{A.~Rubio}
\affiliation{Max Planck Institute for the Structure and Dynamics of Matter,
Center for Free Electron Laser Science, 22761 Hamburg, Germany}
 \affiliation{Nano-Bio Spectroscopy Group, Universidad del Pa\'is Vasco, , 20018 San Sebasti\'an, Spain }
\affiliation{Center for Computational Quantum Physics (CCQ), The Flatiron Institute, 162 Fifth Avenue, New York NY 10010}


\title{
Cavity quantum-electrodynamical polaritonically enhanced electron-phonon coupling and its influence on superconductivity
}
\date{\today}
\begin{abstract}
Laser control of solids was so far mainly discussed in the context of strong classical nonlinear light-matter coupling in a pump-probe framework. Here we propose a quantum-electrodynamical setting to address the coupling of a low-dimensional quantum material to quantized electromagnetic fields in quantum cavities. Using a protoypical model system describing FeSe/SrTiO$_3$ with electron-phonon long-range forward scattering, we study how the formation of phonon polaritons at the 2D interface of the material modifies effective couplings and superconducting properties in a Migdal-Eliashberg simulation. We find that through highly polarizable dipolar phonons, large cavity-enhanced electron-phonon couplings are possible but superconductivity is not enhanced for the forward-scattering pairing mechanism due to the interplay between coupling enhancement and mode softening. An analysis of critical temperature dependencies on couplings and mode frequencies suggests that that cavity-enhanced superconductivity is possible for more conventional short-range pairing mechanisms. Our results demonstrate that quantum cavities enable the engineering of fundamental couplings in solids paving the way to unprecedented control of material properties.
\end{abstract}
\maketitle

\section{Introduction}
Strong coupling and manipulation of matter with photons in quantum-electrodynamical (QED) environments is becoming a major research focus across many disciplines. Among the topics with large potential are the creation of exciton-polariton condensates \cite{byrnes_exciton-polariton_2014}, polaritonic chemistry \cite{galego_cavity-induced_2015, ebbesen_hybrid_2016, flick_kohnsham_2015, flick_atoms_2017} and transport \cite{hagenmuller_cavity-assisted_2018}, quantum nanoplasmonics \cite{tame_quantum_2013}, light-induced topology \cite{lindner_floquet_2011, claassen_all-optical_2016, hubener_creating_2017} and magnetism in 2D materials \cite{shin_phonon-driven_2018}, and novel spectroscopies \cite{ruggenthaler_quantum-electrodynamical_2018}. In condensed matter, the search for control knobs that allow to design properties of quantum materials is an ongoing broad research effort \cite{basov_towards_2017}. One possible route is to employ the nonequilibrium dynamics and coherent manipulation of quantum many-body systems with ultrashort laser pulses \cite{rini_control_2007, forst_nonlinear_2011, zhang_dynamics_2014, mitrano_possible_2016, knap_dynamical_2016, sentef_theory_2017, pomarico_enhanced_2017, kennes_transient_2017, sentef_light-enhanced_2017, tancogne-dejean_ultrafast_2017}. However, in these cases ``classical'' light was typically used.  Here we propose a new route towards manipulating microscopic couplings in solids and inducing ordered phases especially at interfaces and in two-dimensional materials. 

The discovery of enhanced superconductivity in monolayer FeSe on SrTiO$_3$ \cite{liu_electronic_2012, qing-yan_interface-induced_2012, huang_monolayer_2017} and its possible relation to a cross-interfacial electron-phonon coupling \cite{lee_interfacial_2014, rademaker_enhanced_2016, zhang_role_2016} has stimulated considerable interest with an ongoing open debate \cite{zhou_dipolar_2017, wang_phonon_2017, song_phonon-enhanced_2017, jandke_unconventional_2017, lee_what_2015, kulic_electronphonon_2017}. Irrespective of the outcome of this debate, the interfacial phonon mode under consideration is of particular interest for light-control purposes as it has a dipole moment implying bilinear phonon-photon coupling, while at the same time the phonon also couples bilinearly to in-plane FeSe electrons, with a vertex that is strongly peaked for small momentum transfers known as forward scattering. This combination of features is due to the high degree of anisotropy owing to the interfacial structure. Here we employ a prototypical model system, related to FeSe/SrTiO$_3$, for such extreme forward scattering to investigate how photon-phonon coupling in cavities can affect electron-phonon coupling and phonon-mediated superconductivity.

\newpage
\clearpage
\section{Results}

\subsection{Setup: Two-dimensional material inside a cavity}

In Figure \ref{fig:setup}A we show the setup for a two-dimensional material inside a QED cavity environment with perfectly reflecting mirrors. The mirrors confine the photon modes inside the cavity and can lead to strong light-matter coupling even when only vacuum of the electromagnetic field is considered \cite{todorov_ultrastrong_2010,todorov_few-electron_2014}. Specifically, we propose a layered structure of a 2D material (e.g., monolayer FeSe) on a dielectric substrate with large dielectric constant (e.g., SrTiO$_3$) that further helps confine the cavity photon modes of interest. 

For the particular example of FeSe/SrTiO$_3$, the effect of the cavity is to couple the electromagnetic field of the photons polarized along the $z$ direction, perpendicular to the interfacial plane, to a cross-interfacial phonon mode. Importantly, here we go beyond the often-employed rotating-wave and dipole approximations for the light-matter interaction and use full minimal dipolar coupling including the $J \cdot A$ and $A^2$ terms (see Section B of the Supplementary Materials), which makes the theory manifestly gauge-invariant and avoids unphysical divergences. The phonon has a dipole moment along $z$ that involves motion of the O and Ti ions in the topmost layer of SrTiO$_3$, spatially very close to the FeSe monolayer. Specifically, one quasi-dispersionless optical Fuchs-Kliewer phonon at 92 meV \cite{zhang_role_2016} was identified as the most relevant phonon mode that strongly couples to the FeSe electrons both in angle-resolved photoemission \cite{lee_interfacial_2014} and high-resolution electron energy loss spectroscopies \cite{zhang_role_2016}. The influence of screening on this mode is not settled yet, in particular when it comes to phonon linewidths \cite{zhou_dipolar_2017, wang_phonon_2017}. However, the experimental evidence for its influence on electronic properties \cite{lee_interfacial_2014,zhang_role_2016} is definitely present suggesting to use this mode to build a simplified model Hamiltonian to address the impact of reaching strong light-matter coupling on the superconducting behavior of the material. We specifically use a single-band model for the electrons in two spatial dimensions in a partially filled band with filling $n=0.07$ per spin, as previously used to model the relevant electronic structure fitting angle-resolved photoemission data \cite{rademaker_enhanced_2016}. A bilinear electron-phonon scattering is introduced by a coupling vertex $g(\vq)=g_0 \exp(-|\vq|/q_0)$ that is strongly peaked near momentum $\vq=0$ with a coupling range $q_0$. The coupling strength $g_0$ is adjusted to keep a total dimensionless coupling strength $\lambda \approx 0.18$ independent of $q_0$, where $\lambda$ is determined from the effective electronic mass renormalization $m^*/m = 1+\lambda$ in the metallic normal state above the superconducting critical temperature in absence of the cavity coupling. This conservative choice of $\lambda$ is for instance below the value of 0.25 that was given in Ref.~\onlinecite{zhang_role_2016}.

Through phonon-photon coupling we study phonon-polariton formation in this setting. In Figure \ref{fig:setup}B we show schematically the resulting polariton branches that stem from a gauge-invariant coupling involving both $J \cdot A$ and $A^2$ terms, where $J$ is the current of phononic dipoles associated to an infrared-active phonon mode, and $A$ the electromagnetic gauge field of the photons. The relevant effective coupling strength between photons and phonons is given by the phononic plasma frequency $\omega_{\text{P}} = \sqrt{\frac{4 \pi e^2}{M \nu_{0, 2D} L_z}}$, with $M$ the reduced mass of the phonon (see Section B of the Supplementary Materials). For the 2D system in the cavity the plasma frequency is controlled by the length of the vacuum inside the cavity in $z$ direction, $L_z$, and the 2D unit cell area $\nu_{0, 2D}=L_x L_y / N_x N_y $, with $L_i$, $N_i$ the length and number of unit cells of the system in $i$ direction, respectively. The plasma frequency sets the splitting between the upper and lower polariton branches, reminiscent of the LO-TO splitting in bulk semiconductors. Obviously this splitting is only relevant at very small momenta $q$ since the photon energies become large compared to the phonon frequency quickly as $q$ increases due to the large magnitude of the speed of light. 

The formation of phonon polaritons leads to a redistribution of the electron-phonon coupling vertex into the two polariton branches. In the following, we refer to this coupling between electrons and phonon polaritons as ``electron-phonon coupling'', since the coupling originates from electron-phonon coupling in the free-space setting without cavity, and direct electron-photon coupling is not relevant in our setup. In Figure \ref{fig:setup}C we plot the squares of the coupling vertices between electrons and the respective polaritons as a function of $q/k_F$, where $k_F$ is the Fermi momentum. A realistic value of the coupling range for FeSe/SrTiO$_3$ was estimated as $q_0/k_F \approx 0.1$, as needed to create replica bands in angle-resolved photoemission that duplicate primary band features without significant momentum smearing \cite{lee_interfacial_2014,rademaker_enhanced_2016}. In a microscopic model, this value depends on the distance $h_0$ between the topmost TiO$_2$ layer and the FeSe monolayer as well as the anisotropy of in-plane and perpendicular dielectric constants via $q_0^{-1} = h_0 \sqrt{\epsilon_{\parallel}/\epsilon_{\bot}}$, with realistic estimates $\epsilon_{\parallel}/\epsilon_{\bot} \approx 100$ and $1/(h_0 k_F) \approx 1$. This coupling range is larger than the momentum at which photon and phonon branches cross and mix most strongly in the polariton formation process. This implies that the modification of electron-phonon coupling due to the cavity only happens at very small momenta typically smaller than $q_0/k_F$. Thus to investigate how the degree of forward scattering influences the way in which cavity coupling is able to modify the electronic properties, we employ different values for $q_0/k_F$ below, envisioning that cavity effects are enhanced when $q_0/k_F$ becomes smaller, which would in practice be achieved by making the dielectric-constant anisotropy ratio larger. In Table \ref{table} we summarize the relevant parameter values of the bare material used in our simulations.

\begin{table}
\begin{ruledtabular}
\begin{tabular}{c | c | c | c}
Parameter set & A & B & C \\ \hline
Phonon frequency $\Omega$ [eV] & 0.092 & 0.092 & 0.092 \\ 
Electron-phonon coupling $g_0$ [eV] & 2.25 & 4.455 & 11.1 \\
Coupling range $q_0/k_F$ & 0.105 & 0.053 & 0.026 \\
Dimensionless coupling strength $\lambda$ at 116.5 K & 0.180 & 0.180 & 0.180 \\
\end{tabular}
\end{ruledtabular}
\caption{\label{table} Parameters of the bare material system without the cavity used for the simulations discussed in the main text.}
\end{table}

\subsection{Cavity-enhanced electron-phonon interaction}
The critical question to answer here is how the redistribution of the coupling vertex to the upper and lower polariton branches affect the electronic properties. We investigate this by a diagrammatic approach employing Matsubara Green's functions. We adopt the same approximations used in Ref.~\onlinecite{rademaker_enhanced_2016} and compute the self-consistent Migdal-Eliashberg diagram with dressed electronic Green's function in Nambu space, allowing us to take into account superconducting order. The central quantity is the electronic self-energy $
  \hat{\Sigma}(\vk,i\omega_n) =
    i\omega_n[1-Z(\vk,i\omega_n)]\hat{\tau}_0 + \chi(\vk,i\omega_n)\hat{\tau}_3 +
    \phi(\vk,i\omega_n)\hat{\tau}_1,
$
written in terms of the Pauli matrices $\hat{\tau}_i$, the effective mass renormalization $Z(\vk,i\omega_n)$, the band dispersion renormalization $\chi(\vk,i\omega_n)$, and the anomalous
self-energy $\phi(\vk,i\omega_n)$, which vanishes in the normal state. 

\begin{figure}[ht!] 
  \begin{center}
    \includegraphics[width=0.40\columnwidth]{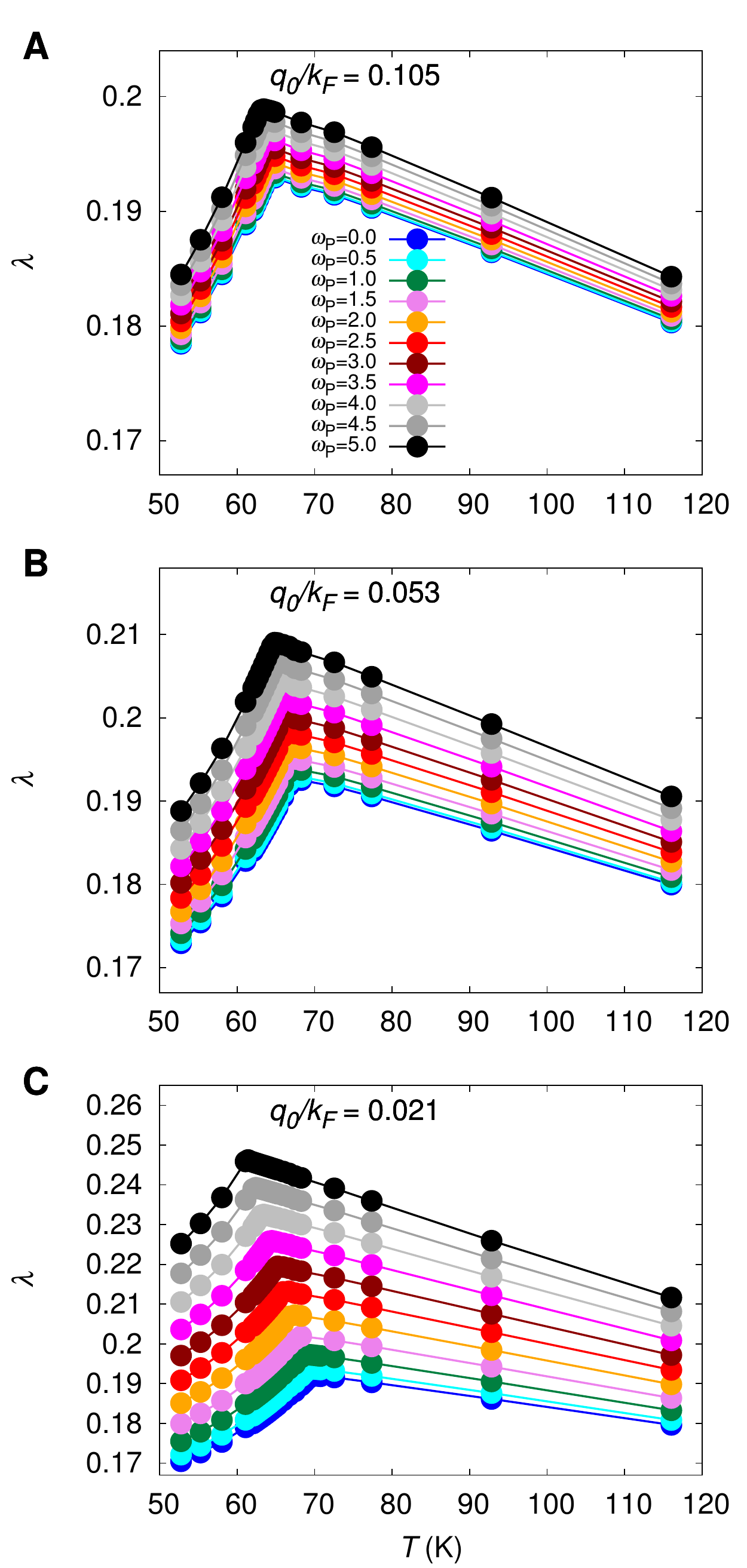}
    \caption{
    \label{fig:lambda} 
{\bf Temperature-dependent electron-phonon coupling for different coupling ranges and plasma frequencies.} 
({\bf A}) The dimensionless electron-phonon coupling strength extracted from the normal self-energy at $\vk_F$ at the smallest Matsubara frequency, $\lambda \equiv Z(\vk_F, i\pi/\beta)-1$, as a function of temperature for a value of the coupling range in momentum space $q_0/k_F = 0.105$ representative of FeSe/SrTiO$_3$, and different phononic plasma frequencies $\omega_{\text{P}}$ as indicated. The case $\omega_{\text{P}}=0$ represents the system without cavity. For increasing $\omega_{\text{P}}$, $\lambda$ increases. Below the superconducting transition, which also shifts with $\omega_{\text{P}}$ (see Figure \ref{fig:delta}), $\lambda$ decreases consistently for all values of $\omega_{\text{P}}$. ({\bf B}) Temperature-dependent $\lambda$ for smaller $q_0/k_F = 0.053$ and different $\omega_{\text{P}}$. As for the superconducting order parameter, the effects of the cavity coupling that is parametrized by $\omega_{\text{P}}$ are more pronounced. ({\bf C}) For even smaller $q_0/k_F = 0.021$, we obtain a strongly enhanced $\lambda$ accompanied by the shift in the superconducting transition that shows up as a cusp in $\lambda(T)$, which reaches a maximum at $T_C$.}
    \end{center}
\end{figure}

We first investigate the effect of the cavity on the effective electron-phonon coupling $\lambda$ itself. This is of interest independently of superconductivity to be discussed below, as the electron-phonon coupling affects also many other properties of materials, such as the conductivity, structural phase transitions, or superconductivity in standard BCS superconductors. In particular it plays a pivotal role for THz-driven nonequilibrium phases of materials. In Figure \ref{fig:lambda} we show how cavity coupling modifies the temperature-dependent quasiparticle mass renormalization obtained from the normal self-energy for the different coupling ranges, realistic $q_0/k_F=0.105$ (Figure \ref{fig:lambda}A), reduced $q_0/k_F=0.053$ (Figure \ref{fig:lambda}B), and very small $q_0/k_F=0.021$ (Figure \ref{fig:lambda}C). The first observation is that independent of the cavity $\lambda$ shows a strong temperature dependence with a peak around $T_C$, decreasing both towards higher temperatures and towards lower temperatures deep inside the ordered phase. The former is readily understood as a usual temperature effect when at high temperature the system becomes more and more classical and less correlated. The latter is understood by considering the fact that correlation effects are reduced deep in the ordered phase when quantum fluctuations lose their importance and a quasi-classical mean-field description can be adopted. Importantly, $\lambda$ is enhanced by the cavity at all temperatures. The cavity effects are more pronounced as $\omega_{\text{P}}$ increases for fixed $q_0/k_F$, and as $q_0/k_F$ increases for fixed $\omega_{\text{P}}$.

\subsection{Light-modified superconductivity}
We now turn to the effect of the cavity on superconductivity. Naively one might expect that an enhanced $\lambda$ leads to enhanced superconducting critical temperature $T_C$. However, the relation is nontrivial as also the effective polariton frequency is relevant for $T_C$. We will see in the following that, unfortunately, for our system the enhancement of $\lambda$ is cancelled by a reduction in the effective frequency.

\begin{figure}[ht!] 
  \begin{center}
    \includegraphics[width=0.45\columnwidth]{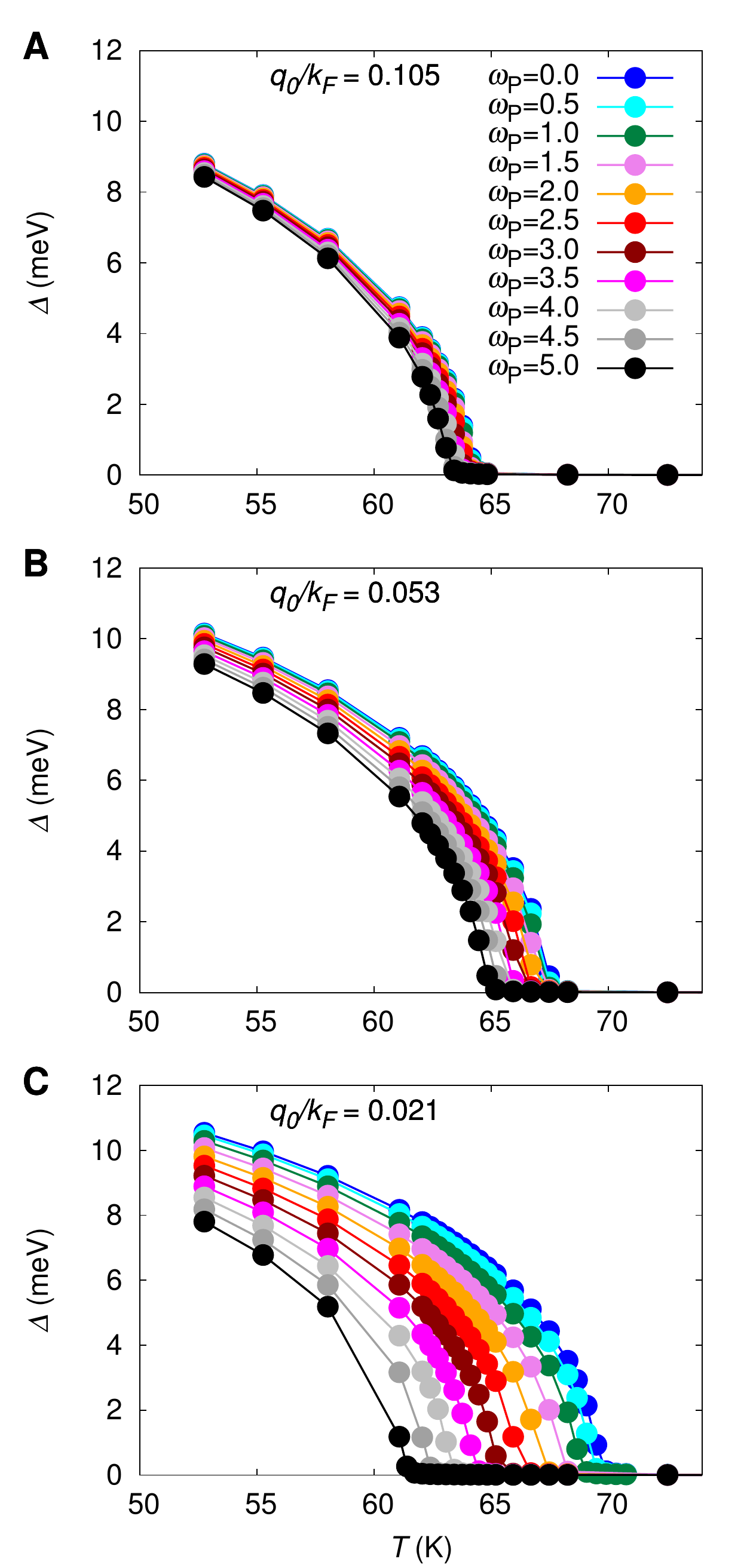}
    \caption{
    \label{fig:delta} 
{\bf Temperature-dependent superconducting gap for different coupling ranges and plasma frequencies.}  ({\bf A}) The superconducting gap at $\vk_F$ at the smallest Matsubara frequency, $\Delta \equiv \phi(\vk_F, i\pi/\beta)/Z(\vk_F, i\pi/\beta)$, as a function of temperature for a value of the coupling range in momentum space $q_0/k_F = 0.105$ representative of FeSe/SrTiO$_3$, and different phononic plasma frequencies $\omega_{\text{P}}$ (measured in eV for the FeSe example) as indicated. The case $\omega_{\text{P}}=0$ represents the system without cavity. For decreasing cavity volume, $\omega_{\text{P}}$ increases, causing a decrease in $\Delta$ and the superconducting critical temperature $T_C$. ({\bf B}) Temperature-dependent gap for smaller $q_0/k_F = 0.053$ and different $\omega_{\text{P}}$. The light-suppressed superconductivity is more pronounced. ({\bf C}) For even smaller $q_0/k_F = 0.021$, strongly reduced $\Delta$ values are observed with increasing $\omega_{\text{P}}$.}
    \end{center}
\end{figure}

Figure \ref{fig:delta}A shows the resulting temperature-dependent superconducting order gap $\Delta \equiv \phi(\vk_F, i\pi/\beta)/Z(\vk_F, i\pi/\beta)$ evaluated at the smallest Matsubara frequency and at a Fermi momentum $\vk_F \approx (0.666/a, 0.666/a)$ along the Brillouin zone diagonal for a coupling range $q_0/k_F = 0.105$ representative of FeSe/SrTiO$_3$. Starting from a critical temperature $T_C \approx 63$ K in the absence of the cavity ($\omega_{\text{P}}=0.0$), we find a slight reduction of superconductivity as the cavity is introduced and its extension $L_z$ in the $z$ direction perpendicular to the 2D material is reduced, resulting in a nonzero $\omega_{\text{P}} \propto 1/\sqrt{L_z}$. For perhaps unrealistically large values $\omega_{\text{P}} = 5.0$ (eV), a reduction of $T_C$ on the order of 1 Kelvin is found in our simulations, which would likely require cavity sizes of a few lattice constants and might in practice be too small to achieve at the moment.

In order to investigate the effect of the forward-scattering coupling range, we look at the change of the superconducting order in the case of $q_0/k_F = 0.053$ that is reduced by a factor of two from the realistic value described above, see Figure \ref{fig:delta}B. In this case the polaritonic redistribution of the coupling is expected to be more effective as there is a better match between the coupling range and the polariton mixing. This is indeed observed in the superconducting order enhancement. Where a value of $\omega_{\text{P}} = 5.0$ was needed in Figure \ref{fig:delta}A to obtain a visible modification of $T_C$, here a smaller value $\omega_{\text{P}} = 2.5$ is sufficient to enhance $T_C$ by $\approx 1$K. Even larger $\omega_{\text{P}}$ lead to enhancements of order 5 $\%$. Finally if we decrease the range by another factor of two, $q_0/k_F=0.021$, the modification is relatively strong with changes of more than 10 $\%$, shifting $T_C$ by up to 10 K (Figure \ref{fig:delta}C).

\subsection{Analysis of the influence of the cavity on superconductivity}
In order to gain physical intuition into why the enhancement of $\lambda$ is insufficient to enhance superconductivity, we take a look at the approximate equation for $T_C$ derived by Rademaker et al.~\cite{rademaker_enhanced_2016} in the extreme forward-scattering and weak-coupling limit:
\begin{align}
T_C &\approx \frac{\lambda \Omega}{2 + 3 \lambda}.
\label{eq:TC}
\end{align}
From this expression it becomes clear that the enhancement of $\lambda$ has to be sufficiently strong compared to the suppression of $\Omega$ that happens concomitantly in our case. This should be contrasted with the standard expression for a momentum-independent coupling vertex in Bardeen-Cooper-Schrieffer (BCS) theory,
$
T_{C, \text{BCS}} \approx 1.13 \Omega \exp(-\frac{1}{\lambda}).
$
The quasi-linearity in $\lambda$ in Eq.~(\ref{eq:TC}) leads to relatively high $T_C$ for moderate values of $\lambda$, but in the cavity also has the negative effect that the enhancement of $T_C$ scales only linearly rather than exponentially with $\lambda$.

\section{Discussion}
Unfortunately, the enhancement of $\lambda$ predicted here does not lead to an enhancement of the superconducting critical temperature $T_C$ in our chosen setting. This effect is explained by the linear scaling of the critical temperature with $\lambda$ for the case of extreme forward scattering in contrast to the exponential scaling for momentum-independent coupling. However, for more conventional pairing mechanisms not geared towards forward scattering, the observed enhancement of $\lambda$ could naturally lead to enhanced $T_C$. Moreover, our theory and the analytical estimates of $T_C$ are valid only in the Migdal-Eliashberg regime of weak coupling, unrenormalized polaritons, and adiabaticity. A polaritonic enhancement of $\lambda$ could still lead to enhancement of $T_C$ even for the forward-scattering case at intermediate couplings, when feedback effects on the polaritons become important, and when nonadiabatic effects come into play. Similarly, interplay between polaritonic pairing and other pairing mechanisms such as spin or orbital fluctuations are subjects for future study. It is possible that in such cases our original motivation of this work, namely to enhance superconductivity in a cavity, might work out. 

In summary we propose to employ QED cavity settings to control polaritonically mediated effects in low-dimensional materials. In reality the size of the achieved effects will depend on the quality factor of the cavity, the degree to which our idealized boundary conditions are realized in practice, and on the required large coupling strengths that can actually be reached in real devices. Importantly, however, our above results are ground-state modifications that are still qualitatively valid even in dissipative systems \cite{de_liberato_virtual_2017,flick_atoms_2017}. Moreover for organic molecules in cavities the ultrastrong-coupling regime was even achieved in bad cavities with small quality factors \cite{bahsoun_electronic_2018}. Here we predict changes of $T_C$ in a few percent range for few-percent changes of the electron-phonon coupling $\lambda$. 
Known examples of LO-TO splitting in bulk semiconductors such as GaP suggest typical ratios of $\omega_{\text{P}}/\Omega$ of order 10 $\%$ \cite{mahan_many-particle_2000}, an order of magnitude smaller than the ones employed in this work. However, we caution that these are very different materials from the ones employed here, and oxide dielectrics close to the ferroelectric phase transition, such as SrTiO$_3$, were suggested to have giant LO-TO splittings exceeding 50$\%$ of the TO frequency \cite{zhong_giant_1994} due to enhanced Born effective charges placing them much closer to the values explored here. It remains to be answered how large realistic LO-TO splittings can become at interfaces. It will definitely be important to explore strategies for enhancing the plasma frequency by synthesizing samples using different substrates with strongly coupled polar phonons, and exploring interface and heterostructure engineering to optimize the dielectric environment.  

We note that a related idea of exciton-mediated superconducting pairing \cite{allender_model_1973} in 2D heterostructures was introduced \cite{laussy_exciton-polariton_2010} and recently discussed in the context of transition-metal dichalcogenides \cite{cotlet_superconductivity_2016}. These proposals require exciton-polariton condensates to exist in the first place, which then affect pairing in doped nearby layers via coupling of quasifree electrons to condensed exciton polaritons. By contrast, our present proposal does not rely on bosonic condensation but rather focusses on directly modifying the electron-phonon coupling through polariton formation in a cavity. For the example of FeSe/SrTiO$_3$, our proposal could help shed light on the above-mentioned debate about the role of the forward-scattering phonon for superconductivity. If the coupling of the phonon to electrons is unimportant, the polaritonic effects will not play a role, which could serve as a test for the influence of the phonon on the electronic properties. Similarly, it was recently suggested to use classical lasers in a pump-probe setting to study the forward-scattering nature of the phonon \cite{kumar_identifying_2017}. Ongoing work focuses on a realistic \textit{ab initio} computation of cavity-enhanced couplings via dipolar phonons using the framework of quantum-electrodynamical density functional theory \cite{ruggenthaler_quantum_2014}.

\textit{Note added in revision.} Upon revision of the manuscript we became aware of two related works, that discuss related ideas of modifying superconducting properties by electron-photon interactions in cavities \cite{Schlawin_cavity-mediated_2018, Curtis_cavity_2018}. 

\section{Materials and Methods}

We employ a cavity quantum-electrodynamical setting with plane-wave mode expansion inside a cavity, with fixed-node boundary conditions for confined cavity photon modes along the $z$ direction, and periodic boundary conditions in the extended 2D plane (see Section A of the Supplementary Materials). Specifically, we use the Migdal-Eliashberg approximation to the electronic self-energy to a coupled electron-polariton model Hamiltonian involving electron-phonon forward scattering and dipolar phonon-photon coupling.

The electron-polariton Hamiltonian has the form 
\begin{align} \nonumber
  H & = \sum_{\vk,\sigma} \epsilon^{\phantom\dag}_\vk c^\dag_{\vk,\sigma}c^{\phantom\dag}_{\vk,\sigma}
    + \frac{1}{\sqrt{N}}\sum_{\vk,\vq,\sigma,\lambda=\pm} c^\dag_{\vk+\vq,\sigma}
    c^{\phantom\dag}_{\vk,\sigma} (g^*_\lambda(\vq) \alpha^\dag_{-\vq,\lambda} + g_\lambda(\vq) \alpha^{\phantom\dag}_{\vq,\lambda}) 
    + \sum_{\vec{q}, \lambda=\pm} \omega_{\lambda}(\vec{q}) \alpha^{\dagger}_{\vec{q},\lambda} \alpha^{}_{\vec{q},\lambda},
\end{align}
with $c^\dag_{\vk,\sigma}$ ($c^{\phantom\dag}_{\vk,\sigma}$) the electron creation (annihilation) operators at
wavevector $\vk$ and spin $\sigma$, $\epsilon_\vk = -2t[\cos(k_x a)+\cos(k_y a)] - \mu$ the electronic band dispersion measured relative to the chemical potential
$\mu$ which is adjusted to fix a band filling of 0.07 per spin. Furthermore $N$ is the number of $k$ points in the 2D Brillouin zone, and $g_\lambda(\vq)$ is the polariton-momentum $\vq$-dependent electron-polariton coupling to branch $\lambda=\pm$,
\begin{align}
g_+(\vq) &= i \sin(\theta_\vq) \sqrt{\frac{\omega_+(\vec{q})}{\Omega}} \; g_0 \exp(-|\vq|/q_0), \\
g_-(\vq) &= i \cos(\theta_\vq) \sqrt{\frac{\omega_-(\vec{q})}{\Omega}} \; g_0 \exp(-|\vq|/q_0), 
\end{align}
with bosonic polariton creation (annihilation) operators $\alpha^{}_{\vec{q},\lambda}$ ($\alpha^{\dagger}_{\vec{q},\lambda}$) for the polaritons with energies
\begin{align}
\omega_{\pm}(\vq) &= \left(\frac12 \left( \omega_{\text{phot}}(\vec{q})^2 + \omega_{\text{P}}^2 + \Omega^2 \pm \sqrt{(\omega_{\text{phot}}(\vec{q})^2 + \omega_{\text{P}}^2 + \Omega^2)^2 - 4 \omega_{\text{phot}}(\vec{q})^2 \Omega^2
 } \right) \right)^{\frac12}.
\end{align}
The unitary transformation from phonons and photons to polaritons is parametrized by
\begin{align}
\arctan(\theta_\vq) &= \frac{\omega_{\text{phot}}(\vec{q})^2 + \omega_{\text{P}}^2 - \Omega^2 + \sqrt{(\omega_{\text{phot}}(\vec{q})^2 + \omega_{\text{P}}^2 + \Omega^2)^2 - 4 \omega_{\text{phot}}(\vec{q})^2 \Omega^2 }}{2 \Omega \omega_{\text{P}}}. 
\end{align}
Here the underlying bare energies are given by the electronic hopping $t = 0.075\ut{eV}$ \cite{rademaker_enhanced_2016}, the phonon frequency $\Omega = 92 \ut{meV}$ \cite{zhang_role_2016}, the bare photon dispersion is $\omega_{\text{phot}}(\vec{q}) = c |\vq|$ with speed of light $c$, and we use a variable effective phononic plasma frequency $\omega_{\text{P}}$ throughout the main text. Further details can be found in Sections B and C of the Supplementary Materials.

The Migdal-Eliashberg electronic self-energy on the Matsubara frequency axis is given by
\begin{align} \nonumber
  \hat{\Sigma}(\vk,i\omega_n) =
    \frac{-1}{N\beta}\sum_{\vq,m,\lambda=\pm} |g_\lambda(\vq)|^2 D_\lambda^{(0)}(\vq,i\omega_n-i\omega_m)
    \hat{\tau}_3\hat{G}(\vk + \vq,i\omega_m)\hat{\tau}_3,
\end{align}
with self-consistent electronic Nambu Green's function $\hat{G}$, decomposed into Pauli matrices $\hat{\tau}_i$, unrenomalized polaritonic Green's function $D^{(0)}$, and fermionic Matsubara frequencies $\omega_n = (2 n+1) \pi /\beta$ and bosonic Matsubara frequencies $\omega_n = 2 n \pi /\beta$, $n \in \mathbb{Z}$, and inverse temperature $\beta = (k_B T)^{-1}$. This amounts to the approximation that the bare phonon mode already contains the energy-shift renormalization due to electron-phonon coupling as the bare phonon frequency is taken from experimental data, and further renormalizations of the phonon polaritons due to electron-polariton coupling are small. The self-consistent computation of $\hat{\Sigma}$ is initialized with a seed for the anomalous superconducting self-energy of $0.007$ eV and a convergence criterion of $10^{-6}$ eV. Further details can be found in Section D of the Supplementary Materials.

\bibliographystyle{Science}
\bibliography{FeSe}

\section{Acknowledgments}
Discussions with H.~Appel, S.~Johnston, S.~Latini, A.~J.~Millis, and L.~Rademaker are gratefully acknowledged. M.A.S. acknowledges financial support by the DFG through the Emmy Noether programme (SE 2558/2-1). A. R. acknowledges financial support by the European Research Council (ERC-2015-AdG-694097), Grupos Consolidados (IT578-13), and  European Union's H2020  program under GA no. 676580 (NOMAD).

\newpage
\section{Figures and tables}
\begin{figure}[ht!] 
  \begin{center}
    \includegraphics[width=0.42\columnwidth]{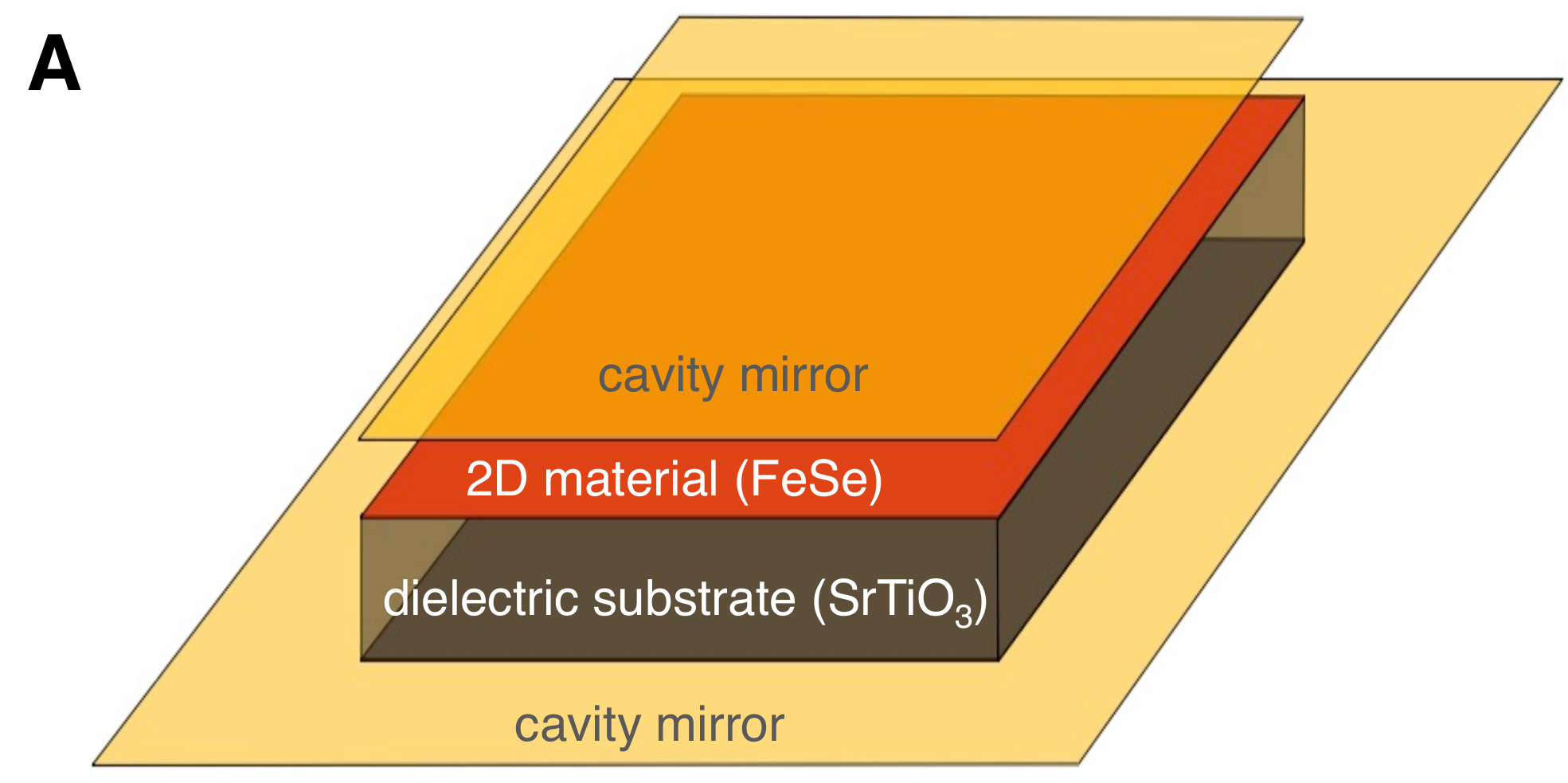} \\
    \includegraphics[width=0.47\columnwidth]{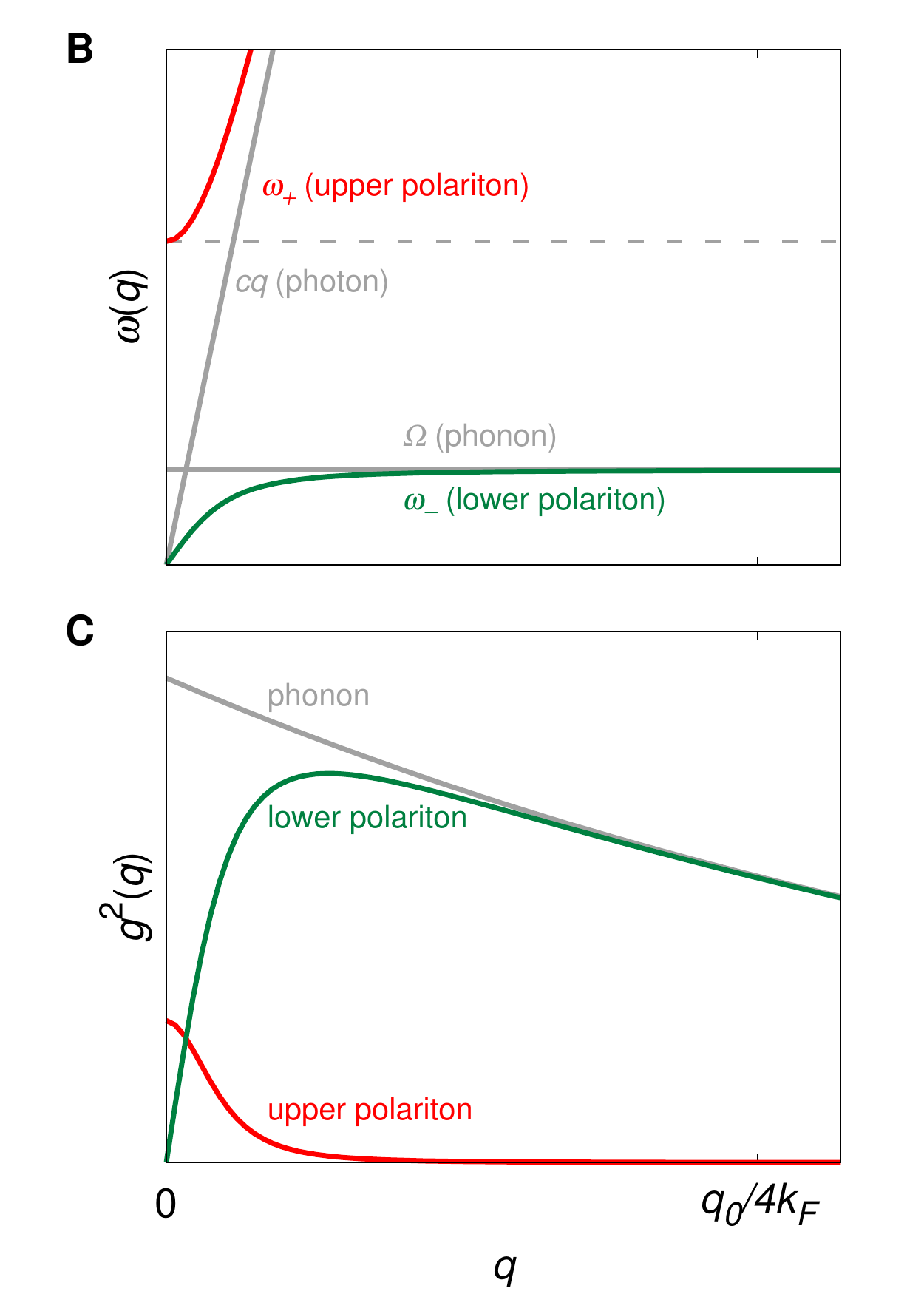}
    \caption{
    \label{fig:setup} 
    {\bf Setup of 2D material in optical cavity, phonon polariton frequency dispersions, and momentum-dependent electron-phonon coupling vertices for the polariton branches.}  ({\bf A}) We consider a setup with a 2D material on a dielectric substrate inside a small optical cavity with mirrors as shown. ({\bf B}) Schematic phonon, photon, upper and lower polariton dispersions versus 2D in-plane momentum $q$. The coupling of the phononic dipole current to the photonic vector potential leads to a splitting given by the plasma frequency $\omega_{\text{P}}$. In the cavity $\omega_{\text{P}}$ is controlled by the cavity volume. ({\bf C}) Momentum-dependent squared electron-boson vertex $g^2(q)$. For forward scattering, the squared bare electron-phonon vertex $g^2(q) = g_0^2 \exp(-2 q/q_0)$ is peaked near $q=0$. In the polaritonic case ($\omega_{\text{P}}>0$) the upper polariton branch inherits some of the electron-phonon coupling at small $q$.}
  \end{center}
\end{figure}

\newpage

\newpage

\clearpage
\newpage

\section{Supplementary materials}

Text (Sections A to D)

References \cite{mahan_many-particle_2000}.

\renewcommand{\theequation}{S\arabic{equation}}

\subsection{Relevant photon modes in cavity}

In this work we consider a 2D material on a dielectric substrate in a nanocavity. We impose reflecting mirror boundary conditions with $\vec{n}\cdot\vec{B}=0$ and $\vec{n}\times\vec{E}=0$ for the magnetic $\vec{B}$ and electric $\vec{E}$ components of the photonic field, and $\vec{n}=\hat{z}$ the surface normal. The size of the cavity in $z$ direction is $L_z$. If the dielectric substrate has a very high dielectric constant, such as for SrTiO$_3$ at low temperature, it can be considered almost metallic and $L_z$ is reduced accordingly in our effective description.

Assuming periodic boundary conditions in the $x-y$ plane, we obtain for example for the vacuum electric field, obeying the wave equation $\nabla^2 E - \frac{1}{c^2}\frac{\partial^2 E}{\partial t^2}=0$ with $c$ the speed of light,
\begin{align}
E_x(x,y,z,t) &= E_1 \exp(i k_x x) \exp(i k_y y) \sin(k_z z) \exp(-i \omega_{\text{phot}}(\vec{k}) t), \\
E_y(x,y,z,t) &= E_2 \exp(i k_x x) \exp(i k_y y) \sin(k_z z) \exp(-i \omega_{\text{phot}}(\vec{k}) t), \\
E_z(x,y,z,t) &= E_3 \exp(i k_x x) \exp(i k_y y) \cos(k_z z) \exp(-i \omega_{\text{phot}}(\vec{k}) t),
\end{align}
with $\omega_{\text{phot}}({\vec{k}}) = c |\vec{k}|$, and 
\begin{align}
k_x &= \frac{2 \pi l}{L_x}, \; l \in \mathbb{N}_0\\
k_y &= \frac{2 \pi m}{L_y}, \; m \in \mathbb{N}_0\\
k_z &= \frac{\pi n}{L_z} \; n \in \mathbb{N}_0.
\end{align}
We assume $L_x$ and $L_y$ to be large to obtain a fine momentum grid in the $x-y$ plane. By contrast $L_z$ is assumed to be small ($L_z \ll L_x, L_y$), implying that for $n = 1$ the photon energy is at least $c \frac{\pi}{L_z}$ well above typical phonon energy scales and thus irrelevant to the problem of our interest. We retain only the $n=0$, $k_z=0$ component that has constant mode amplitude along the $z$ direction. Thus we will use only one mode for each in-plane momentum $\vec{q}=(q_x,q_y)$ with
\begin{align}
E_x(x,y,z,t) &= 0,  \\
E_y(x,y,z,t) &= 0, \\
E_z(x,y,z,t) &= E_3 \exp(i q_x x) \exp(i q_y y) \exp(-i \omega_{\text{phot}}(\vec{k}) t).
\end{align}

\subsection{Phonon-photon Hamiltonian}

We consider the generic Hamiltonian for phonon-photon coupling \cite{mahan_many-particle_2000},
\begin{align}
H_{\text{phon-phot}} &= H_0 + H', \\
H_0 &= \Omega \sum_{\vec{q}} b^{\dagger}_{\vec{q}} b^{}_{\vec{q}} + \sum_{\vec{q}} \omega_{\text{phot}}({\vec{q}}) a^{\dagger}_{\vec{q}} a^{}_{\vec{q}}, \\
H' &= -\frac{e}{Mc} \sum_j \vec{P}_j \cdot \vec{A}(\vec{R}_j) + \frac{e^2}{2Mc^2} \sum_j \vec{A}(\vec{R}_j) \cdot \vec{A}(\vec{R}_j).
\end{align}
Throughout we approximate the phonon dispersion relevant for FeSe/SrTiO$_3$ with a dispersionless $\Omega = 92 \ut{meV}$ \cite{zhang_role_2016}. Here $\vec{q}$ summations are over the first Brillouin zone $[-\pi,\pi)^2$ in the 2D square lattice with lattice constant $a=1$, implying a high-frequency cutoff to the photons, which is irrelevant to the electron-boson physics happening at much lower energy. For the photon, we take only the mode polarized along the $\hat{z}$ direction parallel to the phonon dipoles, and restrict it to the lowest branch $q_z=0$ due to cavity confinement as discussed above, implying $\omega_{\text{phot}}({\vec{q}}) = c |\vec{q}| = c \sqrt{q_x^2+q_y^2}$.

We write the phononic dipole current operator via bosonic operators
\begin{align}
\vec{J}_j &\equiv \frac{e}{M} \vec{P}_j = i e \sum_{\vec{q}} \left(\frac{\Omega}{2 N M} \right)^{1/2} \hat{\xi}_{\vec{q}} \left(b^{\dagger}_{\vec{q}} - b^{}_{-{\vec{q}}} \right) e^{-i {\vec{q}} \vec{R}_j} \equiv \sum_{\vec{q}} \frac{1}{\sqrt{N}} \vec{J}(\vec{q}) e^{-i {\vec{q}} \vec{R}_j},
\end{align}
with polarization vector $\hat{\xi}_{\vec{q}} = \hat{z}$, and similarly for the relevant $z$ component of the photonic vector potential 
\begin{align}
A_z (\vec{R}_j) &\equiv \sum_{\vec{q}} \left(\frac{2 \pi c^2}{\omega_{\text{phot}}({\vec{q}}) \nu_0} \right)^{1/2} \left(a^{\dagger}_{\vec{q}} + a^{}_{-{\vec{q}}} \right) e^{-i {\vec{q}} \vec{R}_j} \equiv \sum_{\vec{q}} \frac{c}{\sqrt{\nu_0}} A_\mu(\vec{q}) e^{-i {\vec{q}} \vec{R}_j},
\label{eq:vector}
\end{align}
assuming periodic boundary conditions inside the 2D plane. Here $b^\dag_\vq$ ($b^{\phantom\dag}_\vq$) creates (annihilates) a phonon with
wavevector $\vq$; $a^\dag_\vq$ ($a^{\phantom\dag}_\vq$) creates (annihilates) a cavity photon with wavevector $\vq$. $N$ is the number of unit cells, $V$ the system volume, $\nu_0 \equiv V/N$ the unit cell volume, and $e$ and $M$ the ionic charge and reduced mass, respectively, related to the relative motion of positively and negatively charged ions in the optical phonon mode. In momentum space we have
\begin{align}
J_z(\vec{q}) \equiv i e \left(\frac{\Omega}{2 M} \right)^{1/2} \left(b^{\dagger}_{\vec{q}} - b^{}_{-{\vec{q}}} \right), \\
A_z (\vec{q}) \equiv \left(\frac{2 \pi}{\omega_{\text{phot}}({\vec{q}})} \right)^{1/2} \left(a^{\dagger}_{\vec{q}} + a^{}_{-{\vec{q}}} \right).
\end{align}

Now we first diagonalize the bare photon plus $A^2$ terms of the Hamiltonian,
\begin{align}
H_{0,\text{phot}} &= \sum_{\vec{q}} \omega_{\text{phot}}({\vec{q}}) a^{\dagger}_{\vec{q}} a^{}_{\vec{q}} \\
&= \frac12 \sum_{\vec{q}} \left(P_{A,\vec{q}} P_{A,-\vec{q}} + \omega_{\text{phot}}(\vec{q})^2 X_{A,\vec{q}} X_{A,-\vec{q}}\right), \\
H_{A^2} &= \frac12 \sum_{\vec{q}} \omega_{\text{P}}^2 X_{A,\vec{q}} X_{A,-\vec{q}},
\end{align}
Here we introduced canonical position and momentum operators for photon degrees of freedom,
\begin{align}
X_{A,\vec{q}} &\equiv \sqrt{\frac{1}{2 \omega_{\text{phot}}(\vec{q}) }} \left(a^{}_{\vec{q}} + a^{\dagger}_{-\vec{q}}\right), \\
P_{A,\vec{q}} &\equiv -i \sqrt{\frac{\omega_{\text{phot}}(\vec{q})}{2}} \left(a^{}_{-\vec{q}} - a^{\dagger}_{\vec{q}}\right).
\end{align}
We also defined the phononic plasma frequency
\begin{align}
\omega_{\text{P}} &\equiv \sqrt{\frac{4 \pi e^2}{M \nu_0}} = \sqrt{\frac{4 \pi e^2}{M \nu_{0, 2D} L_z}},
\end{align}
which for the 2D system in the cavity is governed by the length of the vacuum inside the cavity in $z$ direction, $L_z$, and the 2D unit cell area $\nu_{0, 2D}$. The expressions above are given in cgs units. In the SI system, $\omega^{\text{SI}}_{\text{P}} = \sqrt{\frac{e^2}{M \epsilon_0 \nu_{0, 2D} L_z}}$ with the vacuum permittivity $\epsilon_0$.


The bilinear $J \cdot A$ coupling term is written as 
\begin{align}
H_{J \cdot A} &= - \frac{1}{\sqrt{\nu_0}} \sum_{\vec{q}} \vec{J}(\vec{q}) \cdot \vec{A}(-\vec{q}) \\
&= -\sum_{\vec{q}} \omega_{\text{P}} X_{A,\vec{q}} P_{B,\vec{q}},
\end{align}
where it is convenient to introduce canonical position and momentum operators for the phonons,
\begin{align}
X_{B,\vec{q}} &\equiv \sqrt{\frac{1}{2 \Omega }} \left(b^{}_{\vec{q}} + b^{\dagger}_{-\vec{q}}\right), \\
P_{B,\vec{q}} &\equiv -i \sqrt{\frac{\Omega}{2}} \left(b^{}_{-\vec{q}} - b^{\dagger}_{\vec{q}}\right).
\end{align}
Written in these operators, the bare phonon term $H_{0,\text{phon}} \equiv \Omega \sum_{\vec{q}} b^{\dagger}_{\vec{q}} b^{}_{\vec{q}}$ takes the form
\begin{align}
H_{0,\text{phon}} &= \frac12 \sum_{\vec{q}} \left(P_{B,\vec{q}} P_{B,-\vec{q}} + \Omega^2 X_{B,\vec{q}} X_{B,-\vec{q}}\right).
\end{align}

The total phonon-photon Hamiltonian is now written as pairs of coupled harmonic oscillators,
\begin{align}
H_{\text{phon-phot}} &= H_{0,\text{phot}} + H_{0,\text{phon}} + H_{A^2} + H_{J \cdot A} \\
&= \frac12 \sum_\vq \Big( P_{A,\vec{q}} P_{A,-\vec{q}} + P_{B,\vec{q}} P_{B,-\vec{q}} + (\omega_{\text{phot}}(\vec{q})^2 + \omega_{\text{P}}^2) X_{A,\vec{q}} X_{A,-\vec{q}} \;+ \nonumber \\
&\;\;\;\;\;\;\;\;\;\;\; + \Omega^2 X_{B,\vec{q}} X_{B,-\vec{q}} - 2 \omega_{\text{P}} X_{A,\vec{q}} P_{B,\vec{q}} \Big).
\end{align}
In order to diagonalize this Hamiltonian, we introduce a transformation
\begin{align}
\tilde{P}_{B,\vq} &\equiv \Omega X_{B,\vq}, \\
\tilde{X}_{B,\vq} &\equiv -\Omega^{-1} P_{B,\vq},
\end{align}
which leaves the canonical commutator unchanged but interchanges position and momentum operators. 
The phonon-photon Hamiltonian is then compactly represented as
\begin{align}
H_{\text{phon-phot}} 
&= \frac12 \sum_{\vec{q}} \left[
\begin{array}{c}
P_{A,\vec{q}}\\
\tilde{P}_{B,\vec{q}}
\end{array} 
\right]^T
\left[
\begin{array}{cc}
1 & 0 \\
0 & 1 
\end{array}
\right]
\left[
\begin{array}{c}
P_{A,-\vec{q}}\\
\tilde{P}_{B,-\vec{q}}
\end{array} 
\right] + \nonumber \\
& \;\;\;\; + \frac12 \sum_{\vec{q}} \left[
\begin{array}{c}
X_{A,\vec{q}}\\
\tilde{X}_{B,\vec{q}}
\end{array} 
\right]^T
\left[
\begin{array}{cc}
\omega_{\text{phot}}(\vec{q})^2 + \omega_{\text{P}}^2 & \Omega \omega_{\text{P}} \\
\Omega \omega_{\text{P}} & \Omega^2 
\end{array}
\right]
\left[
\begin{array}{c}
X_{A,-\vec{q}}\\
\tilde{X}_{B,-\vec{q}}
\end{array} 
\right].
\end{align}
Diagonalization is now achieved with the following unitary transformation to polariton canonical position and momentum operators,
\begin{align}
\left[
\begin{array}{c}
X_{+,\vec{q}}\\
X_{-,\vec{q}}
\end{array} 
\right] &= \left[
\begin{array}{cc}
\cos(\theta_\vq) & \sin(\theta_\vq) \\
-\sin(\theta_\vq) & \cos(\theta_\vq)
\end{array}
\right]
\left[
\begin{array}{c}
X_{A,\vec{q}}\\
\tilde{X}_{B,\vec{q}}
\end{array} 
\right],
\\
\left[
\begin{array}{c}
P_{+,\vec{q}}\\
P_{-,\vec{q}}
\end{array} 
\right] &= \left[
\begin{array}{cc}
\cos(\theta_\vq) & \sin(\theta_\vq) \\
-\sin(\theta_\vq) & \cos(\theta_\vq)
\end{array}
\right]
\left[
\begin{array}{c}
P_{A,\vec{q}}\\
\tilde{P}_{B,\vec{q}}
\end{array} 
\right],
\end{align}
which leaves canonical commutation relations intact. The resulting phonon-photon Hamiltonian expressed in polaritonic operators is 
\begin{align}
H_{\text{phon-phot}} 
&= \frac12 \sum_{\vec{q}} \left[
\begin{array}{c}
P_{+,\vec{q}}\\
P_{-,\vec{q}}
\end{array} 
\right]^T
\left[
\begin{array}{cc}
1 & 0 \\
0 & 1 
\end{array}
\right]
\left[
\begin{array}{c}
P_{+,-\vec{q}}\\
P_{-,-\vec{q}}
\end{array} 
\right] + \nonumber\\
& \;\;\;\; + \frac12 \sum_{\vec{q}} \left[
\begin{array}{c}
X_{+,\vec{q}}\\
X_{-,\vec{q}}
\end{array} 
\right]^T
\left[
\begin{array}{cc}
\omega_+(\vq)^2 & 0 \\
0 & \omega_-(\vq)^2
\end{array}
\right]
\left[
\begin{array}{c}
X_{+,-\vec{q}}\\
X_{-,-\vec{q}}
\end{array} 
\right],
\end{align}
with polaritonic dispersions $\omega_\pm(\vq)$ fulfilling
\begin{align}
\omega_{\pm}(\vq)^2 &= \frac12 \left( \omega_{\text{phot}}(\vec{q})^2 + \omega_{\text{P}}^2 + \Omega^2 \pm \sqrt{(\omega_{\text{phot}}(\vec{q})^2 + \omega_{\text{P}}^2 + \Omega^2)^2 - 4 \omega_{\text{phot}}(\vec{q})^2 \Omega^2
 } \right). \label{eq:dispersions}
\end{align}
In particular, in the long-wavelength limit one obtains
\begin{align}
\omega_+(\vec{q} \rightarrow 0) &\rightarrow \sqrt{\Omega^2 + \omega_{\text{P}}^2 }, \\
\omega_-(\vec{q} \rightarrow 0) &\rightarrow 0,
\end{align}
as shown for the semiclassical polariton dispersions in Mahan \cite{mahan_many-particle_2000}.
The diagonalization condition is given by 
\begin{align}
\arctan(\theta_\vq) &= \frac{\omega_{\text{phot}}(\vec{q})^2 + \omega_{\text{P}}^2 - \Omega^2 + \sqrt{(\omega_{\text{phot}}(\vec{q})^2 + \omega_{\text{P}}^2 + \Omega^2)^2 - 4 \omega_{\text{phot}}(\vec{q})^2 \Omega^2 }}{2 \Omega \omega_{\text{P}}}. \label{eq:unitary}
\end{align}
Defining bosonic operators for the upper ($\lambda=+$) and lower ($\lambda=-$) polariton branches,
\begin{align}
X_{\lambda,\vec{q}} &\equiv \sqrt{\frac{1}{2 \omega_{\lambda}(\vq) }} \left(\alpha^{}_{\vec{q},\lambda} + \alpha^{\dagger}_{-\vec{q},\lambda}\right), \\
P_{\lambda,\vec{q}} &\equiv -i \sqrt{\frac{\omega_{\lambda}(\vq)}{2}} \left(\alpha^{}_{-\vec{q},\lambda} - \alpha^{\dagger}_{\vec{q},\lambda}\right).
\end{align}
we rewrite the phonon-photon Hamiltonian in a very compact polaritonic form:
\begin{align}
H_{\text{phon-phot}} &= \sum_{\vq,\lambda=\pm} \omega_{\lambda}(\vq) \alpha^{\dagger}_{\vq,\lambda} \;\alpha^{}_{\vq,\lambda}.
\end{align}

The transformation from the initial phononic degrees of freedom to the final polaritonic ones is then given by
\begin{align}
X_{B,\vec{q}} &= \frac{1}{\Omega} (\sin(\theta_\vq) P_{+,\vec{q}} + \cos(\theta_\vq) P_{-,\vec{q}}).
\end{align}
For the bosonic operators, this implies 
\begin{align}
b^{\phantom\dag}_\vq + b^\dag_{-\vq} &= -i \sin(\theta_\vq) \sqrt{\frac{\omega_+(\vec{q})}{\Omega}} (\alpha^{}_{-\vq,+} -  \alpha^{\dagger}_{\vq,+}) -i  \cos(\theta_\vq) \sqrt{\frac{\omega_-(\vec{q})}{\Omega}} (\alpha^{}_{-\vq,-} -  \alpha^{\dagger}_{\vq,-})), \label{eq:fulltrafo}
\end{align}
which will give the transformation from electron-phonon to electron-polariton coupling in the following.

\subsection{Electron-polariton Hamiltonian}

The electron-polariton model Hamiltonian for FeSe/SrTiO$_3$ inside the cavity reads
\begin{align} \label{eq:ham}
  H &= H_{e-\text{phon}} + H_{\text{phon-phot}}, \\
  \label{eq:hambare}
  H_{e-\text{phon}} &= \sum_{\vk,\sigma} \epsilon^{\phantom\dag}_\vk c^\dag_{\vk,\sigma}c^{\phantom\dag}_{\vk,\sigma}
    + \frac{1}{\sqrt{N}}\sum_{\vk,\vq,\sigma} g(\vk,\vq)c^\dag_{\vk+\vq,\sigma}
    c^{\phantom\dag}_{\vk,\sigma} (b^{\phantom\dag}_\vq + b^\dag_{-\vq} ).
\end{align}
Here, $c^\dag_{\vk,\sigma}$ ($c^{\phantom\dag}_{\vk,\sigma}$) creates (annihilates) an electron with
wavevector $\vk$ and spin $\sigma$; $\epsilon_\vk$ is the electronic band dispersion measured relative to the chemical potential
$\mu$; $g(\vk,\vq)$ is the momentum dependent electron-phonon coupling. The direct electron-photon coupling of electrons in the FeSe plane to the photon branch of interest is neglected, which amounts to the assumption that the paramagnetic electronic current density $\vec{j}$ inside the FeSe layer is perfectly two-dimensional, thus not coupling to the photonic vector potential $\vec{A}$ which points perpendicular to the plane, implying $\vec{j} \cdot \vec{A} \approx 0$.

Adopting the FeSe/SrTiO$_3$ single-band model from Rademaker \textit{et al.} \cite{rademaker_enhanced_2016}, we take an electronic band dispersion $\epsilon_\vk = -2t[\cos(k_x a)+\cos(k_y a)] - \mu$, where $a$ is the in-plane lattice constant. We set $t = 0.075\ut{eV}$ and use as an initial guess $\mu = -0.235\ut{eV}$, which is adjusted during the self-consistent calculations (see below) to a fixed band filling $n_\uparrow = n_\downarrow = 0.07$ for each spin. We neglect the fermion momentum dependence in the electron-phonon coupling $g(\vk,\vq)=g(\vq)$,
where $\vq$ is the momentum transfer, and use $g(\vq)=g_0 \exp(-|\vq|/q_0)$. Here, $g_0$ is adjusted to fix the total
dimensionless coupling strength $\lambda \approx 0.18$ of the electron-phonon interaction in absence of the cavity coupling, and $q_0$ sets the range of the interaction in momentum
space. 

The electron-polariton expressed in polaritonic bosonic operators is obtained via Eq.~(\ref{eq:fulltrafo}) as
\begin{align} \label{eq:hamfull} 
  H & = \sum_{\vk,\sigma} \epsilon^{\phantom\dag}_\vk c^\dag_{\vk,\sigma}c^{\phantom\dag}_{\vk,\sigma}
    + \frac{1}{\sqrt{N}}\sum_{\vk,\vq,\sigma,\lambda=\pm} c^\dag_{\vk+\vq,\sigma}
    c^{\phantom\dag}_{\vk,\sigma} (g^*_\lambda(\vq) \alpha^\dag_{-\vq,\lambda} + g_\lambda(\vq) \alpha^{\phantom\dag}_{\vq,\lambda}) 
    + \sum_{\vec{q}, \lambda=\pm} \omega_{\lambda}(\vec{q}) \alpha^{\dagger}_{\vec{q},\lambda} \alpha^{}_{\vec{q},\lambda},
\end{align}
where 
\begin{align}
g_+(\vq) &= i \sin(\theta_\vq) \sqrt{\frac{\omega_+(\vec{q})}{\Omega}} \; g_0 \exp(-|\vq|/q_0), \label{eq:couplingplus}\\
g_-(\vq) &= i \cos(\theta_\vq) \sqrt{\frac{\omega_-(\vec{q})}{\Omega}} \; g_0 \exp(-|\vq|/q_0). \label{eq:couplingminus}
\end{align}
The couplings are thus fully determined through Eqs.~(\ref{eq:couplingplus}, \ref{eq:couplingminus}) in connection with Eqs.~(\ref{eq:dispersions}) and (\ref{eq:unitary}).
The polariton branches and couplings to the electrons are shown in Fig.~\ref{fig:setup} in the main text.

\subsection{Migdal-Eliashberg simulations}

The electronic self-energy in Migdal-Eliashberg theory on the Matsubara frequency axis employing Nambu notation reads \cite{rademaker_enhanced_2016}
\begin{align}
  \hat{\Sigma}(\vk,i\omega_n) =
    i\omega_n[1-Z(\vk,i\omega_n)]\hat{\tau}_0 + \chi(\vk,i\omega_n)\hat{\tau}_3 +
    \phi(\vk,i\omega_n)\hat{\tau}_1,
\end{align}
where $\hat{\tau}_i$ are the Pauli matrices, $Z(\vk,i\omega_n)$ and $\chi(\vk,i\omega_n)$ renormalize
the electronic single-particle mass and band dispersion, respectively, and $\phi(\vk,i\omega_n)$ is the anomalous
self-energy, which vanishes in the normal state. In Migdal-Eliashberg theory, the self-energy corresponding to the Hamiltonian (\ref{eq:hambare}) is computed by
self-consistently evaluating 
\begin{align}
  \hat{\Sigma}(\vk,i\omega_n) =
    \frac{-1}{N\beta}\sum_{\vq,m} |g(\vq)|^2 D^{(0)}(\vq,i\omega_n-i\omega_m)
    \hat{\tau}_3\hat{G}(\vk + \vq,i\omega_m)\hat{\tau}_3,
\label{eq:eliashberg1}
\end{align}
where $D^{(0)}(\vq,i\omega_\nu) = -\frac{2\Omega}{\Omega^2 + \omega_\nu^2}$ is the bare phonon
propagator, $\hat{G}^{-1}(\vk,i\omega_n) = i\omega_n\hat{\tau}_0 - \epsilon_\vk\hat{\tau}_3 -
\hat{\Sigma}(\vk,i\omega_n)$ is the dressed electron propagator, $N$ is number of momentum grid points, and
$\beta=1/(k_{\text{B}}T)$ is the inverse temperature.

Inside the cavity with $\omega_{\text{P}}>0$, these well-known equations are modified to account for the Hamiltonian (\ref{eq:hamfull}) by using polariton branches $\lambda=\pm$ instead of the phonon:
\begin{align}
  \hat{\Sigma}(\vk,i\omega_n) =
    \frac{-1}{N\beta}\sum_{\vq,m,\lambda=\pm} |g_\lambda(\vq)|^2 D_\lambda^{(0)}(\vq,i\omega_n-i\omega_m)
    \hat{\tau}_3\hat{G}(\vk + \vq,i\omega_m)\hat{\tau}_3,
\label{eq:eliashberg2}
\end{align}
where $D_\lambda^{(0)}(\vq,i\omega_\nu) = -\frac{2\omega_\lambda(\vq)}{\omega_\lambda(\vq)^2 + \omega_\nu^2}$ is the bare polariton propagator,

In practice, we use an initial guess of $0.007$ eV for the anomalous self-energy and run the self-consistency until a convergence to better than $10^{-6}$ eV is achieved. The 2D momentum grid to sample the Brillouin zone is chosen as $2000 \times 2000$ and convergence checked by comparing against $4000 \times 4000$ grids in selected cases. For the patch around $q=0$ we avoid the point $q=0$ where the lower polariton branch becomes soft since the corresponding propagator diverges in the static $\omega_\nu=0$ case. Under the $q$ integral this divergence is cured. We therefore apply a $q$ coarse graining by averaging $\frac{1}{N_{\text{small}}} \tilde{\sum}_{q} |g(\vq)|^2 D^{(0)}(\vq,i\omega_\nu)$ over $N_{\text{small}}$ small patches ($\tilde{\sum}_{q}$ is the sum inside the momentum patch around $q=0$), and using this averaged function in lieu of $|g(0)|^2 D^{(0)}(0,i\omega_\nu)$, again checking convergence in the momentum grid. The momentum convolution in Equations (\ref{eq:eliashberg1}) and (\ref{eq:eliashberg2}) is performed by fast Fourier transforms to a real-space grid and products on the real-space grid. The Matsubara cutoff is $0.4$ eV for the frequency summations, and convergence in this cutoff also checked. 


\end{document}